\documentclass[a4paper]{spie}  % use this instead for A4 paper
\usepackage{amsmath,amsfonts,amssymb}

\usepackage[utf8]{inputenc}   % lettere accentate
\usepackage[english]{babel}   % sillabazione
\usepackage[usenames,dvipsnames]{xcolor} % for hyperref
\usepackage[colorlinks=true, allcolors=blue, pdftex]{hyperref}
\usepackage{aas_macros}
\usepackage[free-standing-units=true]{siunitx}
\usepackage{paralist}
\usepackage{graphicx}
\usepackage{float}
\DeclareSIUnit\pixel{px}
\title{Final design of Cerberus, \\ a three-headed instrument for the
  OARPAF telescope}

  \author[a,*]{Davide~Ricci}
  \author[b]{Lorenzo~Cabona}
  \author[a]{Davide~Greggio}
  \author[b]{Matteo~Aliverti}
  \author[c]{Luciano~Nicastro}
  \author[a]{Luigi~Lessio}
  
  \affil[a]{INAF - Osservatorio Astronomico di Padova, Vicolo
    dell'Osservatorio 5, 35122 Padova, Italy.}

  \affil[b]{INAF - Osservatorio Astronomico di Brera, Via E. Bianchi 46,
    23807, Merate (LC), Italy.}

  \affil[c]{INAF - Osservatorio di Astrofisica e Scienza dello Spazio (Italy).}
    
\authorinfo{$^*$Contact information: davide.ricci@inaf.it
}

\begin{document}
\maketitle

%% proceedings precedente
% https://www.spiedigitallibrary.org/conference-proceedings-of-spie/11447/2562058/Cerberus-A-three-headed-instrument-for-the-OARPAF-telescope/10.1117/12.2562058.full

\begin{abstract}
  Cerberus will be a new scientific instrument for the alt-az,
  1m-class OARPAF telescope in Northern Italy. Currently, the
  telescope operates with a CCD camera used for imaging and
  photometry. One of the objectives of the project is to improve this
  observing mode with a tip-tilt lens for image stabilization up to
  10Hz. Moreover, a long-slit spectroscopy at $\rm R \sim 5900$ and an
  optical fiber échelle spectroscopy at $\rm R \sim 9300$ observing
  modes will be included.  These addtional two ``heads'' of Cerberus
  will be exclusively selected by moving flat $45\degree$ mirrors by
  means of a linear stage placed in a new custom interface flange. The
  flange will replace the existing one, recovering the included field
  flattener lens, to ensure optical correction of the imaging
  channel. We present the already procured COTS hardware, the
  opto-mechanical design of the interface flange, the results of the
  Zemax ray tracing and the design of the web-based instrument control
  software. %, and the integration schedule.
\end{abstract}

% Include a list of keywords after the abstract

%%%%%%%%%%%%%%%%%%%%%%%%%%%%%%%%%%%%%%%%%%%%%%%%%%%%%%%%%%%%%%%%
\section{Introduction}
\label{sec:intro}

The Regional Astronomical Observatory of the Antola Park (OARPAF) is a
$80\cm$, F8 Cassegrain alt-az telescope in the
Ligurian Apennine\cite{2021RMxAC..53...14R}. One of the Nashmyth focal
stations, dedicated to scientific observations, is provided by a
flange with a field flattener lens for the correction of optical
aberrations over a $45\arcminute$ unvignetted Field of View (FoV).
Here, a SBIG scientific camera and a related filter wheel are placed.
Scientific activities are carried out using this setup, that provides
a diagonal FoV of $36\arcminute$, and are mainly focused on
exoplanetary transit follow-up programmed by the students of the
nearby University of Genoa, under the coordination of INAF
researchers\cite{2021JATIS...7b5003R}.

Recently, consistent steps in the remotization of the structure have
been made, by means of a custom, multi-layered python code that wraps
telescope, dome, camera, and switches control software in an
integrated web-based ecosystem\cite{2022SPIE12186E..0PR}.  Latest
developments, including a generalization to extend the software
adoption in other same-class telescopes, is presented in this
conference\cite{ricci2024easy}.

To exploit all the didactic potential and improve the scientific
return of the observatory, a new instrument called Cerberus was
proposed\cite{2020SPIE11447E..5JC}.
Cerberus aims to replace the current flange to add two new focal
stations, orthogonal to the original one.
The field flattener is recovered and included in a new flange.
In the space between the telescope de-rotator interface and the field
flattener, a linear motor supporting two folding mirrors is hosted.
In the default position, the mirrors do not intercept the beam, that
crosses the field flattener, a SBIG tip-tilt image stabilization
system, and finally reaches the current filter wheel + camera
setup. This is the first, upgraded, ``imaging/photometry head''.
Otherwise, the linear motor inserts a folding mirror in the optical
path.  This mirror intercepts the beam before the field flattener, and
reflects a $\approx 9.5\arcminute$ field towards the second, new
``long slit spectroscopy head''.  A $\approx 4.5\arcminute$ fraction
of this field, corresponding to the slit lenght (see next Sections),
is effectively used.
Finally, the linear motor can move forward to insert another folding
mirror in the optical path. This mirror reflects a
$\approx 9.5\arcminute$ field in the opposite direction, towards the
third, new ``service head'' of Cerberus.
As default, this last focal station will be equipped with an \'echelle
spectrograph.

This project is thought from the beginning with the aim of embedding,
in a single insrument, hardware that has already been procured by INAF
and by the University of Genoa during several rounds of financements.
Additional hardware has been specifically procured.

%
% In Sect.~\ref{sec:obs} we present in general the observatory and its
% status.
%
% In Sect.~\ref{sec:inst} we show the detectors and the devices
% that will be used to build the instrument, while the design of
% Cerberus will be shown in detail in Sect.~\ref{sec:design}.  In
% Sect.~\ref{sec:aox} we discuss the opportunity to introduce a tip-tilt
% corrector for photometry in the optical path, and in
% Sect.~\ref{sec:net} we give an overview of the control network of the
% instrument in the framework of the observatory network.  Conclusions
% are treated in Sect.~\ref{sec:conc}.

%%%%%%%%%%%%%%%%%%%%%%%%%%%%%%%%%%%%%%%%%%%%%%%%%%%%%%%%%%%%%%%%
\section{Hardware}
\label{sec:inst}

% ------------------------------------------------------------------
\begin{table}[h]
    \centering
    \caption{ Telescope and instrument parameters }
    \begin{tabular}{@{}ll@{}}
\hline
                       \textbf{Parameter}                         & \textbf{Value} \\
\hline
           Transmitted field (towards field flattener)            & $36\arcminute$ diameter \\
                         Reflected field                          & $9.5\arcminute$ diameter \\
Plate scale (from $0.29\arcsecond/\pixel$ on the imaging channel) & $3.22\arcsecond/\mm$ \\
 Focal extraction (from mirror center to reflected focal plane)   & $151\mm$ \\
\hline
\end{tabular}
\end{table}
% ------------------------------------------------------------------

Here we list following the hardware material that has already been
procured by the University of Genoa.

\begin{description}
  
% %%%%%%%%%%%%%%%%%%%%%%%%%%%%%%%%%%%%%%%%%%%%%%%%%%%%%%%%%%%%%%%%
\item[Imaging/photometry head:] the following devices compose the
  ``imaging/photometry head'' of Cerberus:

  \begin{inparaitem}
  \item a SBIG STX-16801 $4096\times 4096$, $9\um$
    pixels array CCD camera with a diagonal $36\arcminute$ FoV at a
    resolution of $0.29\arcsecond/\pixel$; and a FW7-STX 7
    position filter wheel. $UBVRI$ standard Johnson-Cousins + $H\alpha$
    filters + one position without filters are available.
    These two devices compose the imaging system currently operating at
    OARPAF.    
  \item a SBIG STX-Guider device as guiding camera, to be placed in
    front of the filter wheel,
    %composed by a small $45\degree$ mirror
    %to address a part of the off-axis beam to a guiding camera
    and provided with a $0.7\times$ focal reducer. It is used to
    calculate tracking offsets for the telescope and to the following
    tip-tilt lens;
  \item a SBIG AO-X tip-tilt lens, placed in front of the
    STX-Guider, to correct PSF wobbling at a $10\Hz$ frequency.
    It is able to be tilted by up to $\pm 2.4\degree$
    % basing on the centroid position of a tracking star on the
    % guiding camera
    to displace the field on the main camera up to
    $\pm 144\um$, corresponding to
    $\pm 16\pixel$ with an accuracy of $0.14\pixel$.
    Our simulations show that with a $3\arcsecond$ seeing, and an
    exposure time of $0.1\second$, 50\% of the pointings have
    more than one star brighter than $V=13$ to be used as reference for
    the AO-X\cite{2020SPIE11447E..5JC}.
  \end{inparaitem}

  % %%%%%%%%%%%%%%%%%%%%%%%%%%%%%%%%%%%%%%%%%%%%%%%%%%%%%%%%%%%%%%%% 
\item[Long slit spectroscopy head:] the following devices compose the
  long slit channel of Cerberus. This mode requires a flat mirror to
  direct the light on a slit:

  \begin{inparaitem}
  \item LHIRES~III spectrograph with two gratings available:
    $2400l/\mm$ and $1200l/\mm$, that
    can be replaced by hand.
    The part of the light passing through the slit falls on a
    diffraction grating, that can be tilted by a micrometer screw,
    allowing to select a spectral range of
    $\approx138\nm$.
    Then, the light focuses on a CCD camera.  We foresee a
    $R \sim 5800$ resolution in the visible band.
    % ($450$--$750\nm$).
    %
    LHIRES slit length is
    $8\mm$\footnote{\url{https://www.shelyak.com/produit/spectroscope-lhires-iii/?lang=en\#specifications-produit}}
    which covers a $4.5\arcminute$ FoV.    
  \item a SBIG STL-11000M camera with $9\um$ pixels.
    We estimate a maximum observable magnitude of $m_V\approx 10$, for
    a signal-to-noise ratio of $100$ and 1 hour exposure with a
    sampling of $0.034\nm/\pixel$ on the above
    mentioned spectral range.
\end{inparaitem}

% %%%%%%%%%%%%%%%%%%%%%%%%%%%%%%%%%%%%%%%%%%%%%%%%%%%%%%%%%%%%%%%% 
\item[Service head:] the following devices compose the default option
  for the third head of Cerberus. They are selected by a flat folding
  mirror:
  
  \begin{inparaitem}
  \item a FLECHAS\cite{flechas} échelle spectrograph with a
    $15\meter$, optical fiber head with an effective resolving
    power $R\sim 9300$, increased to $R\sim 15000$ with an image
    slicer to suppress scintillation effects.
  \item a ATIK11000M camera with $9\um$ pixels for the FLECHAS,
    providing a maximum of efficiency of $\approx 18\%$ at $555\nm$.  The
    wavelength range of the spectrograph optics on the camera is
    $350$--$850\nm$.
    
  \end{inparaitem}

% %%%%%%%%%%%%%%%%%%%%%%%%%%%%%%%%%%%%%%%%%%%%%%%%%%%%%%%%%%%%%%%%
\item[Head selector:] to embed this material in the Cerberus concept,

  \begin{inparaitem}
  \item a $150\mm$ VT-80 \texttt{62309150} PI linear stage has been
    provided by INAF. It will be used as the ``head selector'' motor;
  \item a \texttt{C-863.12} PI controller for the VT-80 has been
    specifically procured;
  \item two $50\mm$ right-angle prisms provided by Edmund Optics,
    alluminated on the long
    edge\footnote{\url{https://www.edmundoptics.eu/f/right-angle-mirrors/12280/}}
    together with optical bench material for the mounting, complete
    the selector.
  \end{inparaitem}
    
\end{description}

\newpage

%%%%%%%%%%%%%%%%%%%%%%%%%%%%%%%%%%%%%%%%%%%%%%%%%%%%%%%%%%%%%%%%
\section{Optical design}
\label{sec:inst}

% ------------------------------------------------------------------
\begin{figure}[h]
  \centering
  \includegraphics[width=0.66\textwidth]{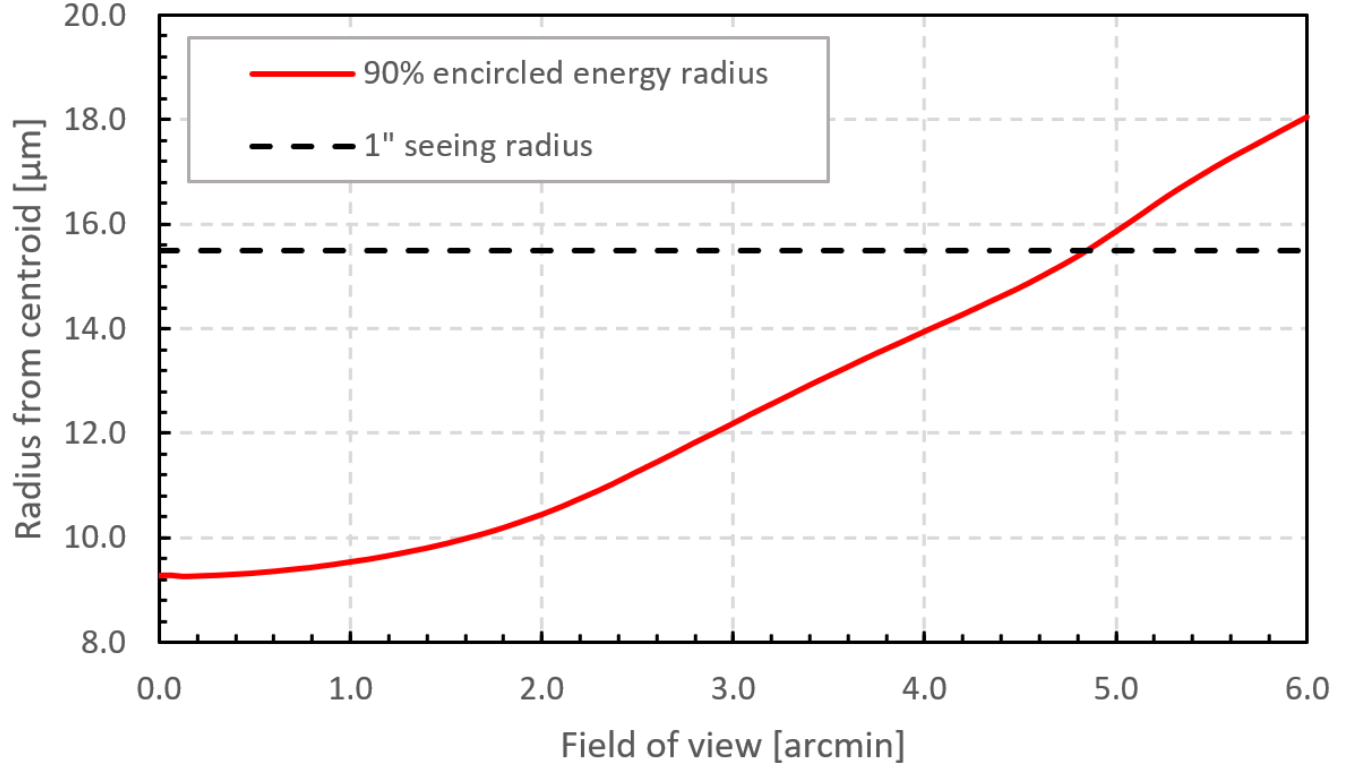}
  \label{fig:fov}
  \caption{ Optical quality on the reflected paths. The Field of View
    is expressed as its radius.}
\end{figure}
% ------------------------------------------------------------------

Given that Cerberus is a project aimed at the maximization of already
procured hardware, we consider its optical design requirements as
a-posteriori.

The first requirement is that the upgrades on the first ``imaging
photometry head'', that introduce an additional backfocus to avoid
mechanical overlapping with the field flattener, can be compensated by
refocusing with the secondary mirror of the telescope without
compromising the optical quality on the diagonal $36\arcminute$ FoV of
the SBIG camera.
This requirement is verified by Zemax simulations after a
reverse-engineering of the field flattener lenses, which specifics are
unknown, except from its mechanical footprint in the existing flange.

The second requirement is that the second ``long slit spectrograph
head'' receives an unvignetted, not aberrated beam on all the $8\mm$
slit length, which covers a $4.5\arcminute$ diameter FOV.
The third and final requirement is that the third ``service head''
allow the largest unvignetted, not aberrated FoV that do not limit
the other ``heads''.

The reflected beams do not pass through the field flattener. The field
aberration, quantified as the radius enclosing $90\%$ of the energy,
is shown in Fig.~\ref{fig:fov}, and compared with the radius
corresponding to a seeing of $1\arcsecond$.  The maximum reflected FoV
is $\approx 9.5\arcminute$ (see Table~1), and corresponds to the point
in which the telescope aberrations are comparable to a very good
seeing conditions at OARPAF.
This allows us to fullfill the second and the third requirement.

This $\approx 9.5\arcminute$ FoV is way more than sufficient to
enlight the $8\mm$-width slit on the ``long slit spectroscopy head'',
and at the same time gives to the ``service head'' a RMS spot radius
without field flattener of $9.4\um$.
To fold this FoV, we purchased two off-the-shelf reflective right
angle, $50\mm$ short edge prisms with coating on the diagonal surface,
in order to simplify the procurement and the alignment on the slit.

These dimensions also let enough stroke to the linear motor to
transmit the $36\arcminute$ FoV when none of the mirrors is in the
optical path.

%%%%%%%%%%%%%%%%%%%%%%%%%%%%%%%%%%%%%%%%%%%%%%%%%%%%%%%%%%%%%%%%
\section{Mechanical design}
\label{sec:inst}

% ------------------------------------------------------------------
\begin{figure}[h]
  \centering
  \includegraphics[width=0.44\textwidth]{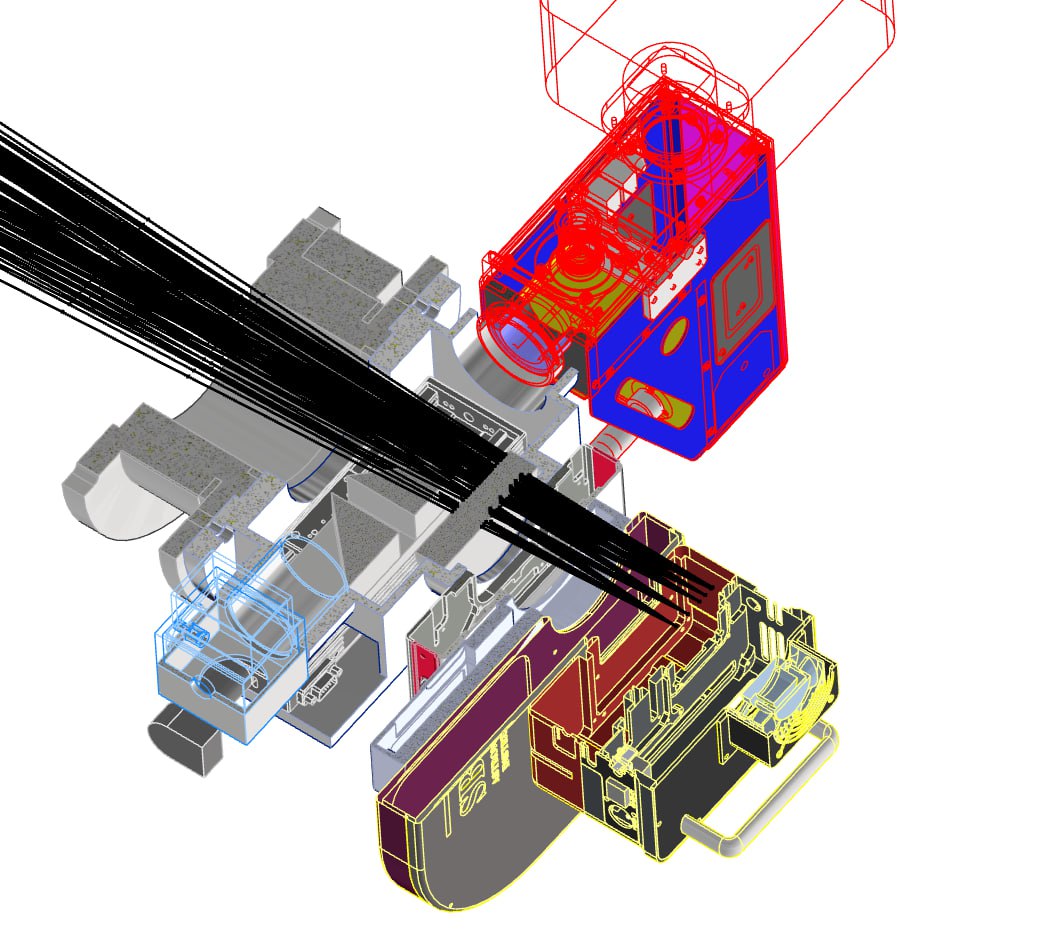}
  \includegraphics[width=0.44\textwidth]{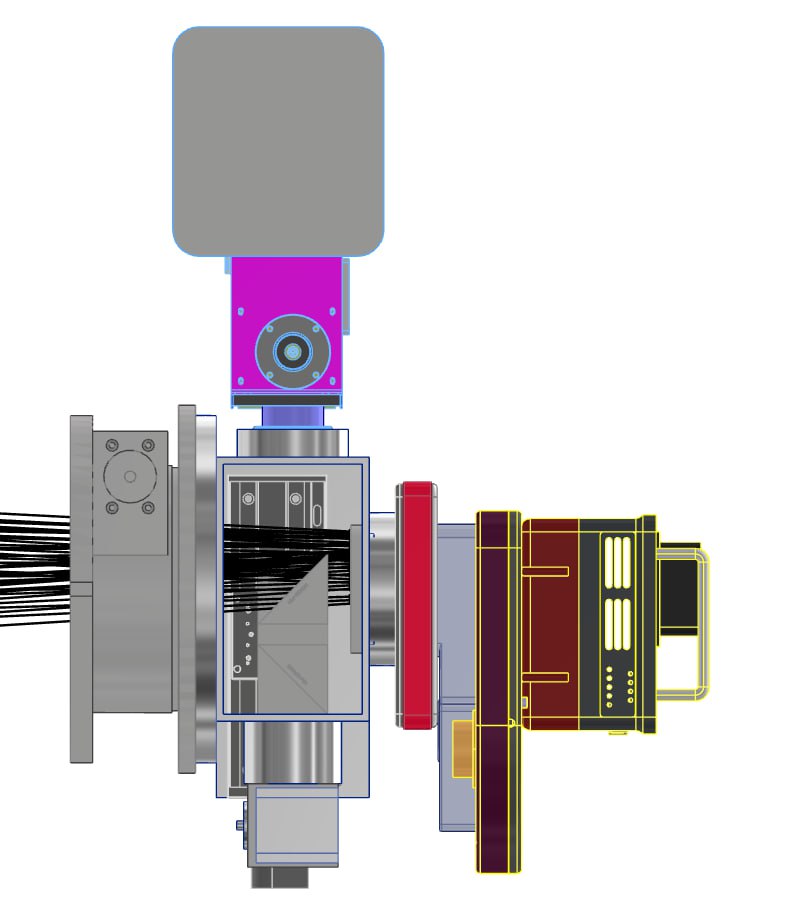}
  \label{fig:mec}
  \caption{ Mechanical design including the PI stage, the right-angle
    reflecting prisms, and the three ``heads'': direct imaging, long
    slit spectroscopy, and the service head initially provided with a
    fiber coupler for échelle spectroscopy.}
\end{figure}
% ------------------------------------------------------------------

An Autodesk Inventor project has been set up to verify mechanical
encumbrances. In particular, VT-80 motor has been placed on the
``Service head'' in order to avoid impacting with the LHIRES~III.

The alluminated right-angle prisms intercept a fraction of the
beam. This allows the external part of the field to reach the
``imaging/photometry head'' in any configuration. Ths can be further
used for secondary guiding / calibration purposes.

%%%%%%%%%%%%%%%%%%%%%%%%%%%%%%%%%%%%%%%%%%%%%%%%%%%%%%%%%%%%%%%%
\section{Software design}
\label{sec:inst}

Cerberus software will be an extension of the existing effort for the
remotization of OARPAF\cite{2022SPIE12186E..0PR}, and that already
provides a user friendly, web-based python environment to control the
telescope, the dome, and power switches.
The current setup also controls the camera + filter wheel, and the
same HTTP API provided by SBIG also allows us to control the AO-X and the
STX-Guider.

STL software will be developed using SBIG SDK, while for ATIK, Alpaca
third-party
drivers\footnote{\url{https://github.com/msproul/AlpacaPi}} can be
adapted.
Finally, preliminary tests on the VT-80 PI stage using the Python
\texttt{pipython} package have been successfully carried out.

While the path of in interfacing with the devices is traced, most of
the effort of the software development is put in defining strategies
to build robust calibration and observation routines. Given the
didatical and scientific goal of the project, students of the
University of Genoa will be involved.

%%%%%%%%%%%%%%%%%%%%%%%%%%%%%%%%%%%%%%%%%%%%%%%%%%%%%%%%%%%%%%%%%%%%%%%
\section{Conclusion}
\label{sec:conc}

We presented the final design of Cerberus, a new interface flange to
embed existing hardware and decectors in an a multi-focal station
instrument for OARPAF: a imaging/photometry station with tip-tilt
corrector, a long slit spectroscopy station, and a service station
initially equipped with an \'echelle spectrograph.

Optical design allowed us to verify the impact of the aberrations and
the feasability, as well as defining the size of the plane mirrors
that will fold the beam into the two new ``heads''.
Mechanical design verified the mechanical footprint of the new
components in a modular layout, so that the ``service head'' can be
easily adapted to other instruments, and the mirrors on the linear
motor do not affect the default ``imaging/photometry'' mode.
The existing control software will be extended to the new components
with the inclusion of new modules.
All additional material opto-mechanical has been procured, and we are
ready to build it!

\acknowledgments DR acknowledges Comitato Interministeriale per la
Programmazione Economica (\texttt{C93C23008400001}).

%%%%%%%%%%%%%%%%%%%%%%%%%%%%%%%%%%%%%%%%%%%%%%%%%%%%%%%%%%%%%%%%%%%%%%%
% References
\bibliographystyle{spiebib} % makes bibtex use spiebib.bst
\bibliography{biblio} % bibliography data in biblio.bib

\begin{thebibliography}{1}

\bibitem{2021RMxAC..53...14R}
{Ricci}, D., {Cabona}, L., {Righi}, C., {La Camera}, A., {Nicolosi}, F., and
  {Tosi}, S., ``{Technical and Software Upgrades Completed and Planned at
  OARPAF},'' in [{\em Revista Mexicana de Astronomia y Astrofisica Conference
  Series}{\nolinebreak\hspace{0.1em}]},  {\em Revista Mexicana de Astronomia y
  Astrofisica Conference Series} {\bf 53},  14--17 (Sept. 2021).

\bibitem{2021JATIS...7b5003R}
{Ricci}, D., {Tosi}, S., {Cabona}, L., {Righi}, C., {La Camera}, A., {Marini},
  A., {Domi}, A., {Santostefano}, M., {Balbi}, E., {Nicolosi}, F., {Ancona},
  M., {Boccacci}, P., {Bracco}, G., {Cardinale}, R., {Dellacasa}, A.,
  {Landoni}, M., {Pallavicini}, M., {Petrolini}, A., {Schiavi}, C.,
  {Zappatore}, S., and {Zerbi}, F.~M., ``{Commissioning and improvements of the
  instrumentation and launch of the scientific exploitation of OARPAF, the
  Regional Astronomical Observatory of the Antola Park},'' {\em Journal of
  Astronomical Telescopes, Instruments, and Systems}~{\bf 7},  025003 (Apr.
  2021).

\bibitem{2022SPIE12186E..0PR}
{Ricci}, D., {Cabona}, L., {Tosi}, S., and {Zappatore}, S., ``{Toward the
  remotization and robotization of the OARPAF Telescope},'' in [{\em
  Observatory Operations: Strategies, Processes, and Systems
  IX}{\nolinebreak\hspace{0.1em}]},  {Adler}, D.~S., {Seaman}, R.~L., and
  {Benn}, C.~R., eds., {\em Society of Photo-Optical Instrumentation Engineers
  (SPIE) Conference Series} {\bf 12186},  121860P (Aug. 2022).

\bibitem{ricci2024easy}
Ricci, D., Cabona, L., Salasnich, B., Nicastro, L., Fini, L., Damonte, A.,
  Tosi, S., and Shibata, T., ``Easy remote observations using web interfaces:
  controlling an italian telescope from japan, and more,'' in [{\em Observatory
  Operations: Strategies, Processes, and Systems
  X}{\nolinebreak\hspace{0.1em}]},   13098--31, SPIE (June 2024).

\bibitem{2020SPIE11447E..5JC}
{Cabona}, L., {Ricci}, D., {Marini}, A., {Santostefano}, M., {Aliverti}, M.,
  {La Camera}, A., {Righi}, C., and {Tosi}, S., ``{Cerberus: A three-headed
  instrument for the OARPAF telescope},'' in [{\em Ground-based and Airborne
  Instrumentation for Astronomy VIII}{\nolinebreak\hspace{0.1em}]},  {Evans},
  C.~J., {Bryant}, J.~J., and {Motohara}, K., eds., {\em Society of
  Photo-Optical Instrumentation Engineers (SPIE) Conference Series} {\bf
  11447},  114475J (Dec. 2020).

\bibitem{flechas}
Mugrauer, M., Avila, G., and Guirao, C., ``Flechas--a new {\'e}chelle
  spectrograph at the university observatory jena,'' {\em Astronomische
  Nachrichten}~{\bf 335}(4),  417--427 (2014).

\end{thebibliography}

\end{document}